# Fast Small-Angle X-ray Scattering Tensor Tomography: An Outlook into Future Applications in Life Sciences

*Christian Appel[a], Margaux Schmeltz[a], Irene Rodriguez-Fernandez[a], Lukas Anschuetz[dg], Leonard C. Nielsen[b], Ezequiel Panepucci[a], Tomislav Marijolovic[a], Klaus Wakonig[a], Aleksandra Ivanovic[acd], Anne Bonnin[a], Filip Leonarski[a], Justyna Wojdyla[a], Takashi Tomizaki[a], Manuel Guizar-Sicairos[af], Kate Smith[a], John H. Beale[a], Wayne Glettig[a], Katherine McAuley[a], Oliver Bunk[a], Meitian Wang[a]\* and Marianne Liebi[abe]\**

a Photon Science Division, Paul Scherrer Institute, Villigen, Switzerland

b Department of Physics, Chalmers University of Technology, Gothenburg, Sweden

c ARTORG Center for Biomedical Engineering Research, Universität Bern, Bern, Switzerland

d Department of Otorhinolaryngology, Head and Neck Surgery, Inselspital, University Hospital and University of Bern, Bern, Switzerland

e Institute of Materials, École Polytechnique Fédérale de Lausanne (EPFL), Lausanne, Switzerland

f Institute of Physics, École Polytechnique Fédérale de Lausanne (EPFL), Lausanne, Switzerland

g Department of Otorhinolaryngology, Head and Neck Surgery, Lausanne University Hospital (CHUV) and University of Lausanne (UNIL), Lausanne, Switzerland

E-mail: meitian.wang@psi.ch, marianne.liebi@psi.ch,

Keywords: saxs, tensor tomography, hierarchically structured materials, nanostrucutral orientation, incus, ossiculoplasty

## Abstract

Small Angle-X-ray Scattering Tensor Tomography (SAS-TT) is a relatively new, but powerful technique for studying the multiscale architecture of hierarchical structures, which is of particular interest for life science applications. Currently, the technique is very demanding on synchrotron beamtime, which limits its applications, especially for cases requiring a statistically relevant amount of sample. This study reports the first SAS-TT measurement at a macromolecular X-ray crystallography beamline, PX-I at the SLS, with an improvement in data acquisition time from 96 h/Mvoxel in the pilot experiments to 6 h/Mvoxel, defining a new standard for fast SAS-TT and allowing the measurement of a full tomogram in 1.2 hours. Measurements were performed on the long and lenticular process of the incus bone, one of the three human auditory ossicles. The main orientation and degree of alignment of the mineralised collagen fibrils are characterised, as well as the size and shape of the mineral particles which show relevant variations in different tissue locations. The study reveals three distinct regions of high fibril alignment, most likely important pathways of sound throughout the ossicular chain, and highlights the potential of the technique to aid in future developments in middle ear reconstructive surgery.

**Introduction**

The development of X-ray diffraction, scattering, and imaging techniques at third-generation synchrotron radiation facilities allowed researchers to visualise materials across multiple length scales. Macromolecular crystallography[1] (MX) has revolutionised our understanding of life at the molecular level by providing three-dimensional structures of biomolecules at atomic resolution (0.1-0.3 nm) and is today widely used by both academia and the pharmaceutical industry. In comparison, Small-Angle X-ray Scattering (SAXS) and Wide-Angle X-ray Scattering (WAXS, also called X-ray diffraction XRD) probe nanoscale features across multiple length scales (0.1 – 200 nm) including hard and soft matter in aqueous solutions or solid forms.[2–6] Scanning the sample with a tightly focused X-ray beam and analysing its diffraction patterns, a technique called scanning SAXS/WAXS or μSAXS/WAXS, allows visualising localised structural heterogeneity across macro-structures in life and materials science applications.[7] Scanning SAXS/WAXS can bridge length scales across macroscopic samples and offers statistical information on nanoscale features volume averaged over the X-ray beam and step size. In Life Sciences, structures with hierarchical order and structures on well-defined length-scales are well suited to be studied with scanning SAXS. Examples include collagen, abundant in the human body, e.g. as part of the extracellular matrix of tissues, myelin sheets wrapping around neurons, or myofilament as a constituent of muscle tissue.[8–14] Their abundance, average orientation and certain structural parameters can be studied using scanning SAXS in 2D or SAXS tensor tomography (SAS-TT) in 3D.[15–17]

SAS-TT was first developed in 2014/2015 at the cSAXS beamline of the Swiss Light Source (SLS) at the Paul-Scherrer-Institute (PSI).[18,19] In SAS-TT experiments, computer tomography (CT) concepts are combined with scanning SAXS and extended to not solely retrieve isotropic intensity values but the full 3D reciprocal space map, from which one can retrieve information about the average nanostructure within each voxel including its orientation. For biomedical research, this 3D nanoscale information can be directly linked to structural properties and provide important structural insights on hierarchically structured materials, for example, the structure of hard bone tissue based on abundance and orientation of calcified collagen or to reveal myelination levels, integrity and axon orientations in a mouse brain.[20–22] While the technique is still under development, multiple synchrotrons (ESRF, MAX IV, PETRA III, DIAMOND, NSLS-II and SSRL4 ) are or have commissioned its use at some of their beamlines. A limiting factor for the technique's application and accessibility to a broader user community is related to time-consuming measurements, which can take more than 24 hrs for a single sample. Such long measurement times are directly related to the technique's capabilities of probing the 3D reciprocal space map of a macroscopic sample, typically millimetres in size. For this purpose, a few hundred 2D projections are taken with a few thousand diffraction patterns each.

To make SAS-TT relevant to a wider user base, particularly for life-science applications in which experiments are required to be performed on a statistically relevant number of samples, the acquisition speed of a full data set needs to be improved significantly. The required experimental setup comprises two translational and two mutually perpendicular rotational motions, as well as a two-dimensional X-ray detector. A typical SAS-TT measurement of a 1.5×2.5 mm$^2$ sample at 25 μm spatial resolution requires approximately ~300 2D projections with 6000 detector images for each 2D projection. For this case, a fully sampled SAS-TT tomogram amounts to 1.8 M detector images, each having typically 2 M to 16 M pixels of about 16-32 bits, in total about 8-20 TB disk usage per dataset which have to be recorded in a reasonable time. A comparative quantity for measurements at different resolutions and sample sizes is the time needed to measure 1 M voxels. The first SAS-TT experiment in 2014 at cSAXS took 35.5 h, corresponding to 96 h/Mvoxel. Since then, multiple beamlines have started to implement the technique and there is a growing demand for considerably reducing acquisition time to boost the applicability and impact of this novel method.

The essential limiting factor for the measurement time per sample is the scanning speed, determined by X-ray beam flux, scanning stages, the frame rate and read-out of the X-ray detector, and their synchronization. The overall increase in flux density at 4$^{th}$ generation synchrotrons will allow faster and finer scanning of samples which in return means however that the deposition rate of the X-ray dose on the sample increases, leading to potential temperature increase and/or resulting in damage of the sample during the measurement. X-ray-induced heating can be mitigated by carrying out measurements under cryogenic conditions, which also significantly reduces X-ray-induced radiation damage, spreading from the atomic to the nanoscale, thus allowing investigations of radiation-sensitive soft materials. The desired instrumentation improvements are met at the macromolecular crystallography beamline PX-I at SLS, which has been optimised for high-throughput cryogenic crystallography including features such as a micro focused X-ray beam with high flux; a multi-axis goniometer - SmarGon for multi-orientation data collection[23,24], a large area and fast readout X-ray detector - EIGER 16M for low-noise data acquisition[25], the implementation of fast 2D continuous scans for serial synchrotron crystallography[26,27], automation from cryogenic sample exchange, software assisted sample alignment with possibility to easily adjust the rotation centre combined with an on-axis optical microscope and suitable online computing resources for fast feedback on the measurements. A Helium filled flight-tube and a semi-transparent beamstop were added to reduce air scattering and record beam transmission for transmission corrections.

In this work, we demonstrate the capabilities and future perspective of SAS-TT at a state-of-the-art macromolecular crystallography (MX) beamline. The measurements reveal the potential for high-throughput SAS-TT experiments, demonstrate for the first time a SAS-TT measurement under cryogenic conditions, and report a record in data acquisition for a full tomogram measured in 1.2 h, respectively at 6 h/Mvoxel, 15 times faster than the first measurements in 2014.

**Results**

The content of this work is structured in two parts for which two samples from the human auditory ossicle, the malleus and incus bone as shown in figure 1B, are used. In the first part, we will focus more on technical aspects related to fast SAS-TT data acquisition and instrumentation aspects. We included a brief study on photon statistics on the malleus to explore how much further fast acquisition schemes can be pushed in the future. In the second part we will present a pilot study on the long and lenticular process of the incus to reveal the potential of SAS-TT as a tool to investigate samples for the life science community.

**Instrumentation & Control System**

The required instrumentation for a SAS-TT setup comprises stages for fast raster scanning, two rotational degrees of freedom, a low-noise large area detector with fast readout capabilities and a control system with optimised motor motion. The state-of-the-art setup at PX-I is very well suited to perform SAS-TT measurements. It is an undulator beamline equipped with a Si(111) double crystal monochromator (DCM) capable of operating at energies between 6 and 20 keV. A two-stage focusing with a dynamically bendable mirror and a Kirkpatrick-Baez (KB) mirror pair allows for variable beam size from 5 times 5 $\mu m^2$ to 80 times 80 $\mu m^2$ at the sample position

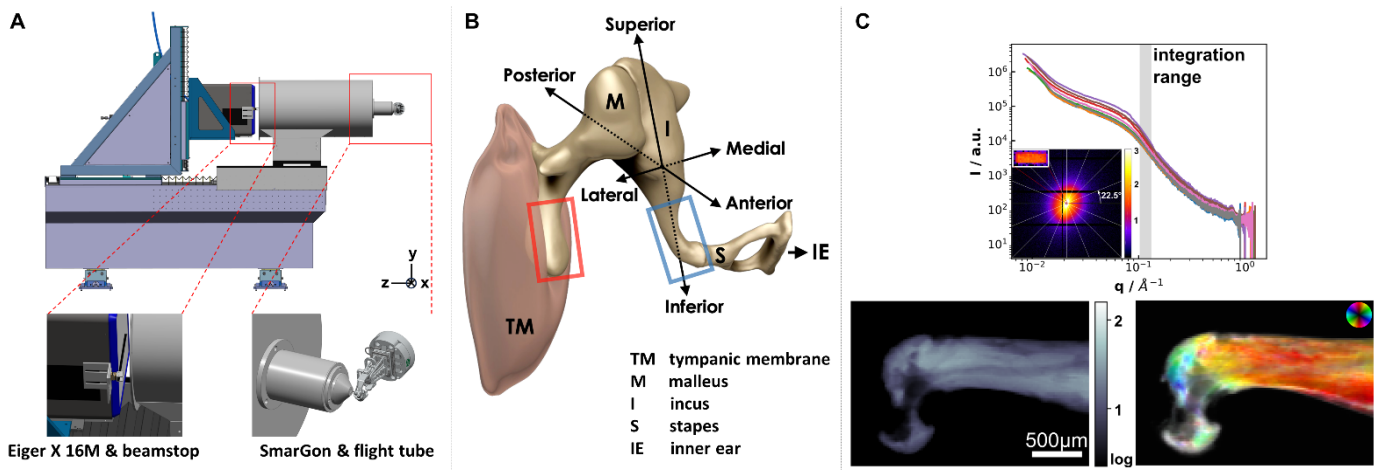

*Figure 1: (A) Drawing of the experimental setup with custom-built flight tube, SmarGon and Eiger X 16M detector together with semi-transparent beamstop. (B) 3D view of the middle ear (adapted from "Radiopaedia - Drawing Middle ear ossicles and tympanic membrane - no labels" at AnatomyTOOL.org by Frank Gaillard, license: CC BY-NC-ND) to show the tympanic membrane and ossicular chain (malleus, incus and stapes). Within the small b/w sketch shown in the inset, the blue and red box highlight the regions that were scanned for the incus and malleus respectively. (C) Segment-wise integrated 1D scattering curves of the incus with an inset of the corresponding 2D scattering pattern, together with 2D absorption and the orientation and abundance plot below. For the orientation and abundance plot, the intensity is integrated within the q range indicated with a grey box in the 1D intensity plot and analysed as described in the main text. Here, hue shows the direction of anisotropy, the value shows the symmetric intensity, and the saturation the degree of orientation.*

28 m from the source. These focusing capabilities determine the spatial resolution of the SAS-TT measurements for which the volume averaged information of the nanoscale can be recorded. Furthermore, PX-I offers some application features designed for macromolecular crystallography under cryogenic conditions which include an automated sample exchange, cryogenic storage of samples and a so-called cryostream for sample cooling during measurement. These features offer opportunities to extend studies on perishable, frozen and radiation sensitive samples as well as fully automated data acquisition schemes for high throughput studies.

The SmarGon (SmarAct GmbH)[24] allows XYZ positioning and three angles of rotation around an arbitrary point in space. It's positioning resolution is < 5 nm. The SmarGon is combined with Aerotech stages with motion and control overheads as little as ~0.4 s in between lines for continuous scanning in *snake motion* (alternatingly forth and back scanning of lines). Therefore, all conditions for optimised acquisition schemes are well met with the given setup. The available range for scanning is 5 times 5 mm², 0…360º and 0…45º for the rotational degrees of freedom ($\varphi$ and $\chi$). The technical rendering in figure 1A shows the setup that facilitates the fast scanning at various orientations of the sample as required for proper sampling of the 3D reciprocal space as needed for the SAS-TT reconstructions. As example for its overall acquisition performance, a sample with dimensions of 2.8 times 1.6 mm² measured at 20 µm resolution and a frame rate of 80 Hz took 16 h at PX-I with 13 h being the total exposure time. This illustrates the highly efficient use of synchrotron light at this beamline.

The beamline's large area detector, the EIGER X 16M, allows acquisition rates up to ~116 Hz in 16 M active pixel mode (32 bit), and rates up to 500 Hz in a reduced area of 4 M active pixels (16 bit). In contrast to regular MX experiments, the detector was placed at 1.2 m distance from the sample to resolve the smaller scattering angles with a custom-built flight tube filled with helium placed in between the sample and detector to reduce air scattering and absorption. The sample-detector distance was calibrated using silver behenate (AgBH) powder as a reference[28]. A 3 times 3.5 mm$^2$ (diameter, thickness) silicon single crystal was installed as a semi-transparent beamstop on a motorised stage in front of the EIGER reducing the intensity of the direct beam to a level that is not saturating the detector, allowing for a simultaneous transmission measurement.

Another factor contributing to the highly efficient use of the synchrotron-radiation beamtime is the on-axis microscope in combination with the graphical user interface (GUI), developed for fast and precise alignment of crystals and the setting up of grid-scans for raster scanning across samples.[26,27] While this is in place at most state-of-the-art MX beamlines, scanning SAXS and SAS-TT beamlines are typically in a much more infant state regarding software and hardware support for sample alignment and this takes correspondingly much longer - on the order of hours rather than minutes at PX-I.

Fast data pipelines with analysis and live-feedback capabilities are crucial for an efficient and well-controlled measurement. We deployed a pipeline adapted from the cSAXS MATLAB package[29] to provide online feedback on data acquisition, and the computing resources at PX-I were capable to keep up with the measurements in real time for all acquisition rates. We computed 16 segments from the raw detector image and reduced them to 8 assuming point-symmetry of the scattering signal. In figure 1C, 1D SAXS profiles are shown for the 8 segments. The scattering signal in each segment can be associated with information about the nanostructure with a specific orientation within the sample. In the case of bone, we use the range of $q=0.5$-$0.6$ nm$^{-1}$ to quantify the content and orientation of the bio mineralised crystals within collagen fibrils of the bone following the procedure described by Bunk et al.[30] The main direction of orientation for the nanoscale features is extracted from the phase of the cosine $\theta$ and the degree of orientation is calculated by dividing the asymmetric intensity $a_1$ (amplitude of the cosine) by the symmetric intensity $a_0$ (average intensity). The orientation and scattering of mineralised collagen can be mapped based on transmission normalised scattering features in each individual pattern and visualised as a pixelated image with help of the colour-coding according to the colour wheel. Here, the hue shows the direction of anisotropy (main orientation), the value shows the symmetric intensity, and additionally, the saturation shows the degree of orientation. Highly oriented material appears with bright colours whereas low-oriented material appears in a tone from black to grey, depending on its abundance.

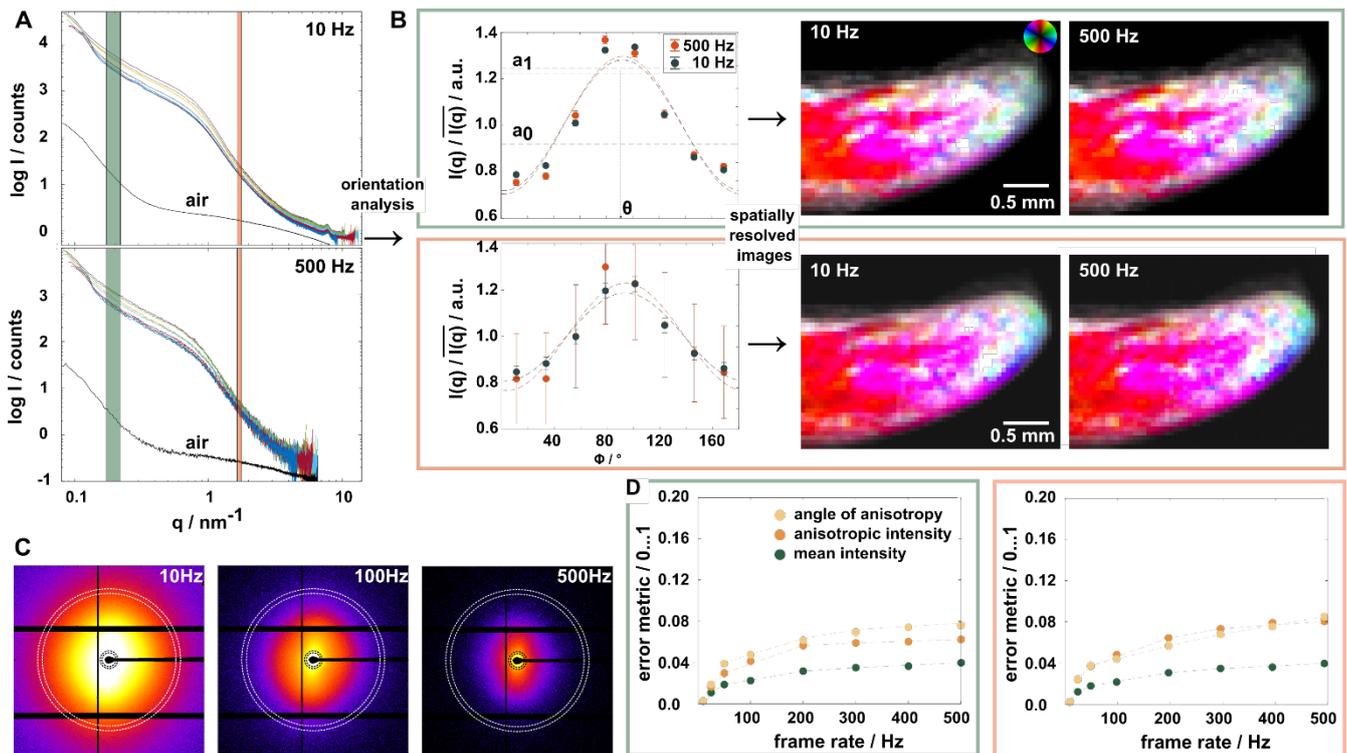

*Figure 2 (A) 1D scattering curves for 8 integrated segments of the diffraction pattern after averaging regularly spaced opposing segments from the EIGER for 10 and 500 Hz, together with air scattering as the background (B) The highlighted q regions in (A) are integrated, and an orientation analysis as described in the text is performed. Error bars are calculated as the standard error of the mean across the diffraction patterns for the respective ranges. The orientation of nano structural features is colour-coded according to the circular legend. (C) 2D scattering patterns are shown for different acquisition rates with two circular dashed double rings that highlight the q ranges used for the orientation analysis. (D) A similarity analysis is performed for both q ranges using all points of the grid scan (error metric described in main text). Dashed lines act as guide for the eye.*

**Perspective for data acquisition at 4$^{th}$ generation synchrotrons**

In recent years, several synchrotrons have been constructed and upgraded to 4$^{th}$ generation such as MAX IV in Sweden, Sirius in Brazil and the ESRF in France, getting much closer to a diffraction limited storage ring (DLRS). The SLS 2.0 upgrade[31,32] started in October 2023, and for the PX-I beamline this includes upgrades for its insertion device and X-ray optics. The beamline is expected to obtain an increase in brilliance by up to two orders of magnitude and potentially 3 orders in terms of overall flux density with a wide bandwidth monochromator. It is unquestionable that a flux-hungry technique such as SAS-TT strongly benefits from this upgrade. However, the increase in flux must be matched with appropriate instrumentation and software for fast data acquisition. We benchmarked the experiment control and instrumentation setup of PX-I with the available flux of 2.56 $10^{12}$ photons/s (calibration performed with glassy carbon standard, GCL14[33]).

The increase in photons and higher dose can also lead to radiation induced damage of the sample, which means that acquisition schemes must be adjusted accordingly. To explore the effect of photon statistics on a relevant sample, we varied the dose on the sample by measuring 2D projections of the malleus from the human ossicular chain with a range of different acquisition rates between 10 and 500 Hz covering a 50-fold increase in dose on the sample. Diffraction patterns of the EIGER at different frame rates are shown in figure 2C. With increasing frame rate, photon counts on the detector reduce which is also visible in the 1D SAXS curves in figure 2A. Measurements above 100 Hz have a reduced q-range. Scattering maps of the orientation of nanostructures from the bone are calculated from the azimuthal plots for two different q-ranges, q = 0.17…0.22 $nm^{-1}$ and q = 1.65…1.77 $nm^{-1}$, to explore low and high photon count statistics from the sample. We choose the integration width in q for both ranges small to be representative for nano structural features with a narrow width, i.e. diffraction like peaks, and to remain comparative to bin sizes used for the q-resolved SAS-TT reconstructions. For 500 Hz, the larger error bars for larger q are related to reduced counting statistics. We can extract the mean intensity $a_0$, anisotropic amplitude $a_1$, and the direction of the anisotropic scattering $\theta$ to present them in the colour-coded images as shown in figure 2B. To compare measurements at different frame rates in a quantitative manner, we use the normalised root-mean-square error (NRMSE) as a measure for the similarity for the mean intensity and the anisotropic amplitude of the images $I(x_i,y_j)$ with the 10 Hz measurement as the reference $R(x_i,y_j)$:

$$NRMSE = \sqrt{1 - \frac{\left(\sum_{i,j} I(x_i, y_j) * R(x_i, y_j)\right)^2}{\left(\sum_{i,j} I(x_i, y_j)\right)^2 * \left(\sum_{i,j} R(x_i, y_j)\right)^2}}$$

For the angle of the anisotropic intensity, namely the main orientation of the nanostructure, we use the square-root of 1 minus the dot product of the image and reference. Results of our error metric are shown in figure 2D, with 0 meaning that two images are identical whereas 1 means that there is no similarity between the images. For all three quantities, we see a slow increase of the error with increasing frame rate. However, the error remains below 8 % even at 500 Hz acquisition. Interestingly, both q ranges give similar results, meaning that the orientation analysis is robust enough against the noise variation in the data. With the upcoming upgrade to SLS 2.0 and the advent of new detector technology such as the Jungfrau[34] or Matterhorn[35] detector, measurements in the 1-10 kHz regime are within reach. Already now, we were able to record a full SAS-TT dataset of the malleus bone at 500 Hz with 150 projections, a field of view of 1.2 times 2 mm² at 25 µm resolution and under cryogenic conditions. Results on the reconstruction are shown in the SI (Figure S1). The full dataset was recorded in a record time of 1.2 hours including all overheads, corresponding to 6 h/Mvoxel.

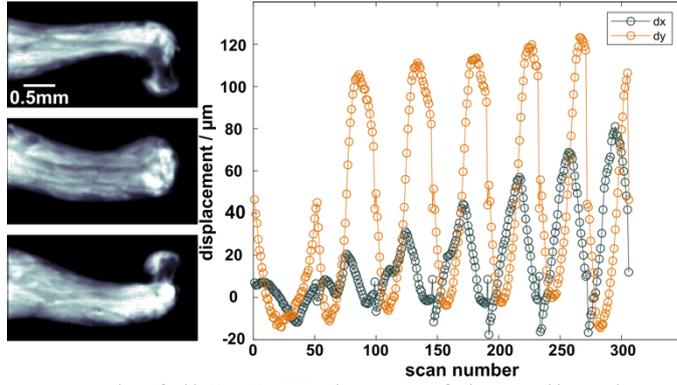

*Figure 3: 2D projections at three different rotation angles are shown for 0 ° tilt. The 1D plot shows the displacement in x and y as a function of the scan number for all tilt and rotation angles. In total, 7 full tomograms were recorded for different rotation angles.*

**SAS-TT measurement: alignment and reconstructions**

A full SAS-TT data acquisition consists of a few hundred 2D projections measured at different rotation and tilt angles, which need to be aligned before reconstruction to correct for any hardware misalignments. In the current case, we use the fully integrated intensity on the detector, commonly known as dark-field signal, to align all projections after the beamtime. We compute a tomogram at 0 tilt using the Filtered Back Projection (FBP) algorithm, and cross correlate all projections with the reconstruction computationally tilted by the same nominal value. The alignment shifts *dx* and *dy* are plotted in figure 3 for the incus. The repetitive motion visible for *dx* and *dy* is related to the 7 tomograms, one for each tilt angle, that were measured chronologically to sufficiently sample the 3D reciprocal space. Shifts that have been determined in this way and corrected for are smaller than 140 µm in total.

Reconstruction of the 3D reciprocal-space map for specific q-ranges is carried out using the software package MUMOTT[36] (*version 0.2.1*) as described in Nielsen et al.[37] Input for the reconstruction are the azimuthal segment-wise integrated intensities for each individual q-range. For the q-resolved reconstruction, we reduce the transmission corrected 1D data into 100 logarithmically equally spaced bins. The 3D reciprocal space in each voxel is reconstructed using band-limited Friedel-symmetric spherical functions expressed in spherical harmonics, with a band-limit of **e**|| = 6. The orientation of the main intensity for each voxel is determined from the eigenvector associated with the smallest eigenvalue of the rank-2 tensor derived from the degree-2 component of the spherical function's polynomial expansion. The robustness of the reconstruction is checked by visual comparison of 2D orientation, anisotropy, and degree of orientation between the measurements and simulated projections of the reconstructed data. The degree of orientation is calculated as the ratio between the standard deviation (anisotropic component) and mean (isotropic component).

**Application Example: Nanoscale survey of a human auditory ossicle**

SAS-TT on hierarchical structures provides unique insights across multiple length scales, as highlighted below with a pilot study on a human incus, one of the three auditory ossicles of the middle ear. The ossicular chain, see Figure 1B, is composed of the malleus, incus and stapes and is responsible for the sound transmission between the external environment and the inner ear, where the sound is transformed to an electrical signal. The impedance difference between the air and the fluid-filled cochlea is matched by the ossicles. Interestingly, studies suggested that the auditory ossicles are already fully developed at birth, with only little bone remodelling happening in a lifespan, in order to preserve the ossicular architecture and ensure a stable sound transmission throughout life.[38] Other studies correlated bone mineral density with sound transmission, showing that a remodelling process lowering the mineral density could happen in regions experiencing more motion and higher forces.[39] It has also been shown that the degree of mineralization and apatite orientation was higher in the ossicles than in long bones.[40] One recent study using SR X-PCI identified a vast vascular network inside the ossicles with associated perivascular zones of less bone density.[41]

Pathologies of the middle ear such as chronic otitis media, otosclerosis or cholesteatoma usually affect ossicle integrity, resulting in hearing loss. These conditions often require reconstructive surgery of the ossicular chain. Therefore, gaining insights on the multi-scale architecture of the ossicles, the orientations of the collagen/mineral components and the location of potential bone-remodelling regions is crucial to improve surgery planning, as well as the design and placement of passive middle ear implants. While a part of the malleus was measured demonstrating fast SAS-TT (SI Figure S1) the long and lenticular processes of the incus were measured with increased photon statistics and analysed in detail as outlined below. With the higher brilliance of 4th generation sources, photon statistics will be of sufficient quality for a full analysis in the fast mode.

**Variation of size and shape of the mineral nanoparticles**

The primary building blocks of bone are collagen fibrils reinforced with small mineral particles of hydroxy-apatite. Amount, size, orientation and distribution of these particles play a pivotal role in defining the mechanical properties of the bone-collagen-mineral composite.[42,43]

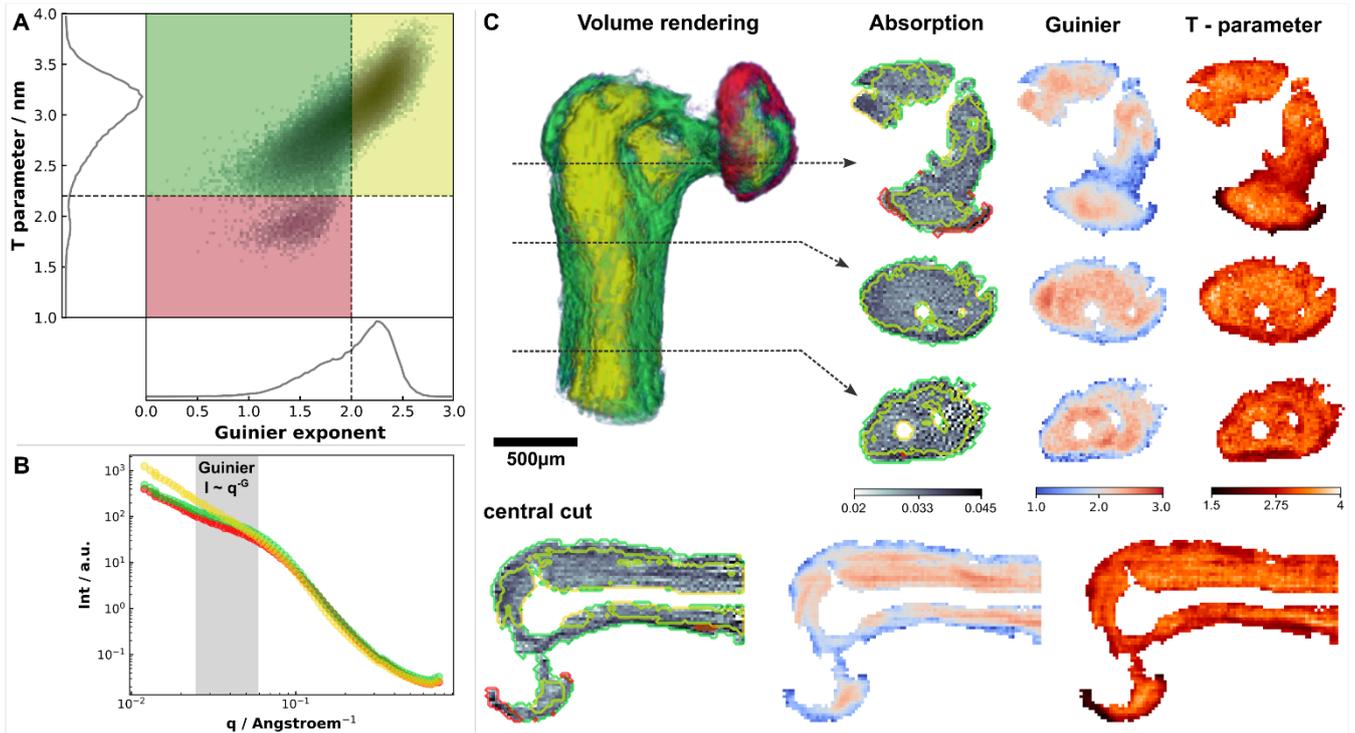

*Figure 4: Q resolved SAXS-TT of the incus. (A) 2D histogram of T parameters and Guinier exponent. Data is segmented based on the histogram in 3 phases (red, green, and yellow), see details on the segmentation in the main text. Representative 1D scattering curves for each phase are shown in (B), with the region used to calculate the Guinier exponent highlighted in grey. In (C), a 3D volume rendering of the segmented long and lenticular processes of the incus is shown as well as slices from different cuts through the volume showing absorption, Guinier exponent and T-parameter. Contour lines colour-coded for the segmented phases are plotted on top of the slice for the absorption contrast.*

Mineralization changes throughout the life cycle of bone[14] as well as within regions where different forces or motions are applied.[39] A low level of mineralization is found in newly formed bone and can be an indicator of sections where bone remodelling occurs within the ossicles. To extract information on the mineral particles, we use the T-parameter model that assumes isotropic scattering from a 2-phase system, a mineral volume fraction of 50 % and sharp interfaces.[44] This model simplifies the calculation of the smallest dimension of the particles to the ratio of the volume V to surface area S: $T = 4\,V/S$. Since the volume to surface ratio depends on the shape of the particles, one typically combines the T-parameter model with computing the dimension of the scattering objects using the Guinier exponent ($I \sim q^{-G}$), where for isolated particles G=1 is characteristic for 1D, i.e. needle like, and G=2 for 2D, i.e. platelet like objects. In solid samples where contributions of form- and structure factor to the scattering the interpretation of the Guinier exponent is less clear, but still can be used to differentiate between shapes of the nanostructure of bone. Surface and volume of the scattering particles can be extracted from the invariant $Q = \int_0^\infty I(q)\,q^2 dq = V/(2\pi^2\,(\Delta\rho)^2)$ and the Porod constant $P = I(q)/q^4 = S/(2\pi\,(\Delta q)^2)$ for a 2-phase system.[43] Here, $\Delta\rho$ depicts the Scattering Length Density (SLD) contrast. The model is fitted to the q resolved reconstruction of the mean intensity to compute the invariant Q from the q range 0.025…0.28 Å$^{-1}$, the Porod constant for 0.16 to 0.28 Å$^{-1}$ and the Guinier exponent for 0.025…0.055 Å$^{-1}$ from the background subtracted data.

The analysis of mineral particles is shown in figure 4. The Guinier exponent and mineral thickness (T-parameter) is computed for all voxels of the 3D reconstructed volume masked with a thresholded absorption tomogram to avoid inclusion of volume outside of the main body. Figure 4A shows the distribution of the T parameter and the Guinier exponent values over all the voxels of the reconstructed volume, in form of a 2D histogram. We note that within this simplified model, a reliable readout of the T-parameter is limited to the region for Guinier exponent ≤ 2, which corresponds to needle (1D) or platelike (2D) particles. In the histogram, we clearly see a population of particles shifted to larger Guinier values. This has been experimentally observed for immature bone, e.g. embryonic long bone mineralization in mice[8] or healing bone close to interface of biodegradable implants.[11] From a scattering point of view this indicates a more compact shape of the mineral particles. Based on the distribution in the 2D histogram, which shows a small secondary side maximum around G~1.7, and the theoretically highest valid G exponent of G=2, we decided to classify our data into three phases based on the following two boundaries of G = 2 and T = 2.2 nm. For each segmented phase, a representative 1D scattering curve is shown in figure 4B, while the spatial distribution of the phases is highlighted in the volume rendering in figure 4C. The segmentation clearly separates the bone into the inner long process (yellow), an interfacial layer surrounding the long process and forming the bony pedicle (green) and an annular thin layer spreading around the lenticular process (red), highlighting the advantage of the multi-contrast images obtained from SAS-TT. Different slices through the 3D volume are shown for three parameters: absorption, Guinier exponent and T parameter. The latter two are nanostructure features extracted from small-angle scattering. The main vascular channel of a branched network known to exist inside the long process of the incus can be identified already in the absorption contrast.[41] These channels are surrounded by the yellow phase with values of the Guinier exponent G > 2. We also observe a correlation with an increased T-parameter, but as mentioned before this calculation is no longer valid due to constraints of the model. The more compact shape of the mineral particles could point towards the hypothesis of bone remodelling happening in this central region, close to the vessels and where the sound pressure is the highest. The phase bordering the long process (green), is more pronounced in the inner lateral part rather than the anterior and posterior sides. The bony pedicle, which is bridging the long process to the lenticular process, consists almost exclusively of this green phase characterised by G < 2 but larger T-parameters. Focussing on the lenticular process, we can see a circular layer in red, which is characterised by significantly lower T-parameters.

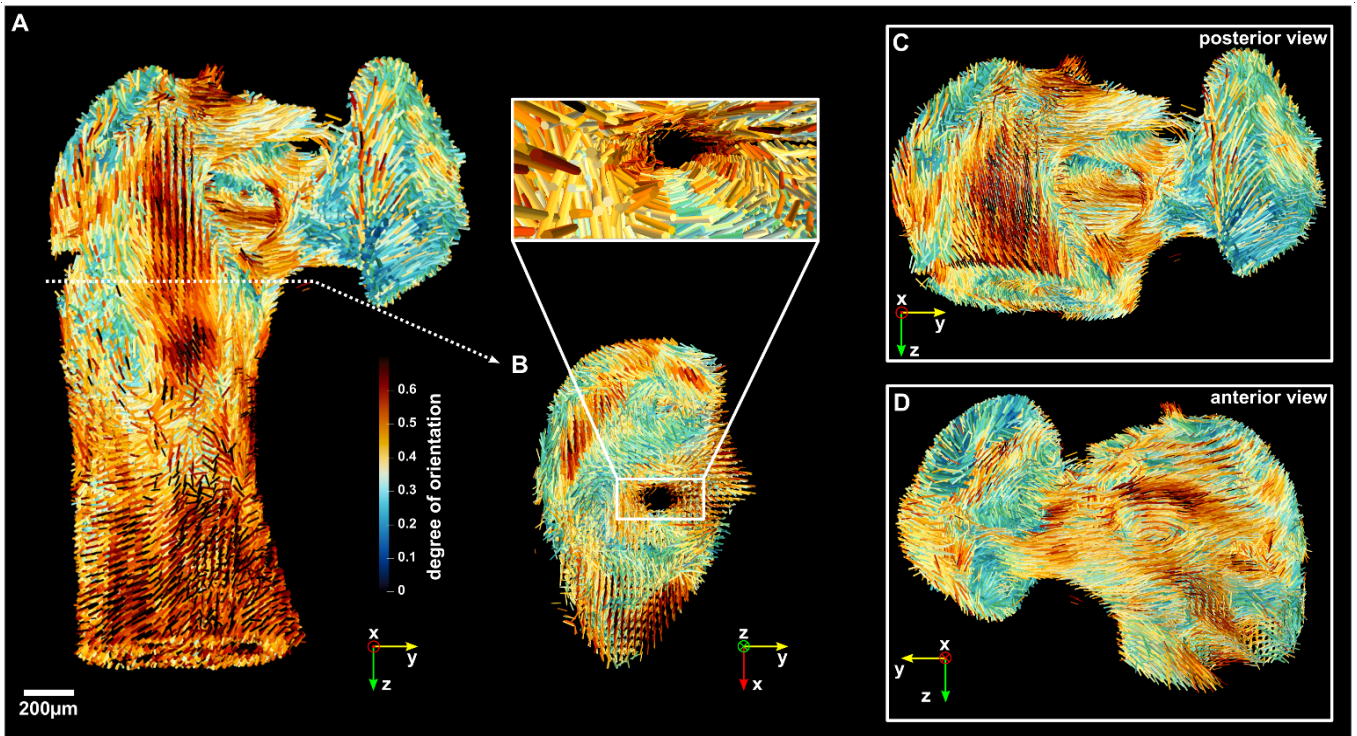

*Figure 5: (A) 3D glyph representation of the orientation of mineralised collagen fibrils within the long and lenticular process of the incus bone derived from SAS-TT (q range 1.61…1.7 nm$^{-1}$). The cylinder size is scaled with the mean intensity while the colour represents the degree of orientation. (C) and (D) highlight the bony pedicle with lenticular processes attached from the posterior and anterior view, while (B) shows a cross-section with a view through the long process. The expanded view along the large vascular channel within the long process illustrates the alignment of collagen fibrils close to the channel.*

**Orientation of mineralised collagen fibrils**

We obtain the orientation of the mineralised collagen fibrils based on the main orientation of the equatorial-like scattering feature in a q range of 1.61…1.7 nm$^{-1}$. Figure 5 illustrates the results in the form of 3D glyph representation with cylinder pointing in the direction of the main structure's orientation contained in one voxel. The size of the cylinders is scaled with the mean intensity of the rank-2 tensor, while the colour represents the degree of orientation calculated from the standard deviation of the same tensor. Figure 5 A-D show different views of the lenticular and long process of the incus using the same 3D glyph representation (coordinate systems attached).

The degree of orientation of the mineralised collagen fibrils seems to be higher in the long process compared to the lenticular process of the incus. The long process has a more regular and elongated shape, as well as an inner vascular network which could be the reason that the mineralised fibrils point into a dominant direction. The region of the bony pedicle, that makes the junction between the 2 processes, particularly visible in figure 5C and 5D (posterior and anterior view), is clinically critical, as it is a non-vascularised region. The bone is nourished by diffusion from the bony pedicle to the lenticular process, which is compromised in case of acute or chronic inflammation. We notice bundles of fibrils highly oriented towards the lenticular process that built three bridges or pair of bridges between the long process and the bony pedicle connecting it to the lenticular process. Towards the lenticular process, the three bridges merge into a single path where fibrils appear to be more randomly oriented again as indicated by the brighter colour.

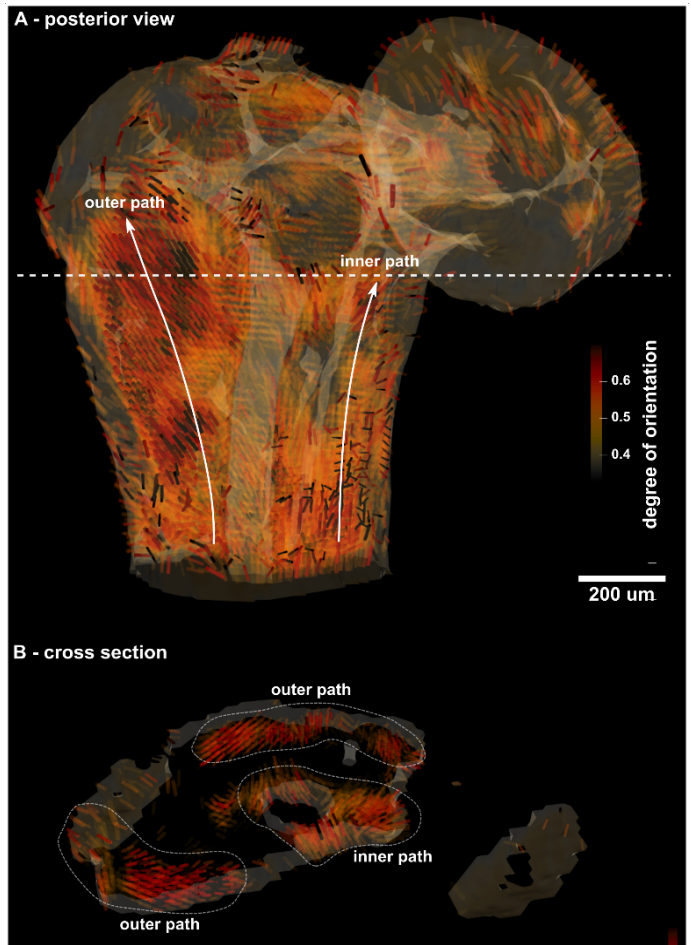

*Figure 6: 3D glyphs superimposed with an isosurface rendering of the absorption tomogram. Fibrils with a degree of orientation below 0.4 are fading out due to a strong transparency mask that is applied. (A) posterior view of the long and lenticular process, the two white lines with an arrow highlight the outer path and inner path extending from the long process to the bony pedicle. (B) shows a cross section extracted from the height highlighted by the dashed white line. The 3 regions with 2 outer and 1 inner path are additionally encircled with dashed lines and tagged.*

Upstream, along the long process, we can observe the existence of 3 main paths showing a high degree of orientation of the mineralised fibrils. They are highlighted in the cross-section shown in figure 5B: two outer paths aligned along the long process, and one wrapping around the main vascular channel, also visible in video V1 of the Supplementary Materials. These three bundles are also illustrated in the posterior view in Figure 6A together with a cross section (6B). Figure 6 is a combination of masking out fibrils of low anisotropy with a strong transparency mask for values < 0.4, superimposed with an isosurface rendering of the absorption tomogram. Video

V2 in the Supplementary Materials gives a better understanding of the 3D arrangement within this view, which supports the existence of these highly oriented paths that bridge the long process and the lenticular process via the bony pedicle. Here, we note that the outer paths appear to be less anisotropic when they wrap around the top of the long process, however, this observation may be attributed to having multiple fibril directions superimposed within single voxels.

Highly anisotropic features within the bone might play a role in the mechanical properties and efficient sound transmission. These findings could be crucial for future surgical planning, as they provide insights into collagen fibril orientation within bone, aiding in reconstructive strategies for the ossicular chain. One concrete example would be the optimization of the attachment points of a passive middle ear implant onto the long process of the incus.

**Discussion**

In this study we show the potential of a state-of-the-art macromolecular X-ray crystallography beamline to perform fast SAS-TT for life science applications. The beamline's setup is very well suited for quick alignment, fast data acquisition and excellent data quality. In addition, the readily available cryogenic conditions may play a pivotal role for radiation sensitive samples. We investigated two samples from the human auditory ossicles, the incus and the malleus. Based on measurements of the malleus at different frame rates, we benchmarked the data quality in terms of different photon statistics and the future potential of SAS-TT that is ready to leverage from the increase in photons at $4^{th}$ generation synchrotron sources. Already with the current photon flux of $2.56\ 10^{12}$ photons/s, we were able to acquire a full SAS-TT dataset with 1.2 times 2.0 $mm^2$ field of view, 25 µm resolution and 150 projections at different rotational degrees of freedom in a record time of 1.2 h. This corresponds to 6 h/Mvoxel which is 15x faster compared to the first SAS-TT measurements in 2015. With the upcoming upgrade program SLS 2.0 and new detector technology, such as the Jungfrau or Matterhorn, data acquisition in the 1-10 kHz range will be in reach.

In the second part of the paper, a pilot study on the long and lenticular processes of the incus highlights the technique's potential impact for studies within the life science community. The analysis focusses on distribution and content of mineral particles, the building unit of mineralised collagen fibrils, as well as their orientation. Collagen organisation is linked to the bone's mechanical properties and can therefore be used as an indicator for understanding the path of sound waves through the bulk of the auditory ossicle. Our analysis reveals that three distinct phases exist in the incus with different shape and size of the mineralised nanoparticles.

We are further able to extract the main orientation of the mineralised collagen fibrils and their anisotropy, which allows us to identify certain regions in the long process that exhibit a higher alignment. Here, we find two outer and one inner path along the largest vascular channel in the centre of the long process. These three paths converge into junctions at the bony pedicle bridging the long process to the lenticular process. Visualisation of these paths together with the bone's absorption gives unique insights into the nano structural arrangement and may be directly utilised in planning and optimisation of the surgery procedure at this critical junction. Given the improvements in terms of acquisition time, future studies may include a larger scope with more samples, which promotes SAS-TT to be a viable technique for more extended studies in life science applications of statistical relevance and give access to a larger user community.

**Author contributions**

ML, OB, and MW conceived the research project. CA, EP, TM, KW, FL, JAW, KS, JHB, WMG, KM, OB, MW, and ML planned and implemented the beamline adaptions. CA, EP, KW, FL, JAW, KS, TT, WMG, KM, MW, and ML carried out the X-ray experiments. CA, IRF, LN, KW, MGS and ML analysed the data. LA, AI, MS and AB provided the sample. CA, MS, IRF, LA, ML interpreted the results. CA, MW and ML drafted the first version of the manuscript. All authors contributed to the final version of the manuscript.

**Acknowledgements**

CA has received funding from the European Union's Horizon 2020 research and innovation program under the Marie Skłodowska-Curie grant agreement No 884104, and is also supported by funding from Chalmers initiative for advancement of neutron and X-ray techniques. LA and AB have received funding from the Swiss National Science Foundation (SNSF) with grant No 320030_192660 and for IRF SNSF grant No 310030E_188993. ML and LN received funding from the European research council (ERC-2020-StG 949301). MS received funding from the PHRT-iPostdoc Project Nr. 2022/477 ORCHESTRAMUS. LN and ML received funding from the Swedish Research Council (VR 2018-041449). We acknowledge the Paul Scherrer Institute, Villigen, Switzerland for provision of synchrotron beamtime at the beamline PX-I of the SLS.

# References

[1] A. Wlodawer, Z. Dauter, M. Jaskolski, Eds., *Protein Crystallography: Methods and Protocols*, Springer New York, New York, NY, **2017**.

[2] R. H. Bragg, M. L. Hammond, J. C. Robinson, P. L. Anderson, *Nature* **1963**, *200*, 555.

[3] K. Wakabayashi, M. Tokunaga, I. Kohno, Y. Sugimoto, T. Hamanaka, Y. Takezawa, T. Wakabayashi, Y. Amemiya, *Science* **1992**, *258*, 443.

[4] M. Povia, J. Herranz, T. Binninger, M. Nachtegaal, A. Diaz, J. Kohlbrecher, D. F. Abbott, B. J. Kim, T. J. Schmidt, *ACS Catalysis* **2018**, *8*, 7000.

[5] C. Appel, B. Kuttich, T. Kraus, B. Stühn, *Nanoscale* **2021**, *13*, 6916.

[6] A. V. Martin, A. Kozlov, P. Berntsen, F. G. Roque, L. Flueckiger, S. Saha, T. L. Greaves, C. E. Conn, A. M. Hawley, T. M. Ryan, B. Abbey, C. Darmanin, *Communications Materials* **2020**, *1*, DOI 10.1038/s43246-020-0044-z.

[7] P. Fratzl, H. F. Jakob, S. Rinnerthaler, P. Roschger, K. Klaushofer, *J Appl Crystallogr* **1997**, *30*, 765.

[8] I. Silva Barreto, S. Le Cann, S. Ahmed, V. Sotiriou, M. J. Turunen, U. Johansson, A. Rodriguez-Fernandez, T. A. Grünewald, M. Liebi, N. C. Nowlan, H. Isaksson, *Adv. Sci.* **2020**, *7*, 2002524.

[9] C. Giannini, L. De Caro, A. Terzi, L. Fusaro, D. Altamura, A. Diaz, R. Lassandro, F. Boccafoschi, O. Bunk, *IUCrJ* **2021**, *8*, 621.

[10] C. Giannini, M. Ladisa, V. Lutz-Bueno, A. Terzi, M. Ramella, L. Fusaro, D. Altamura, D. Siliqi, T. Sibillano, A. Diaz, F. Boccafoschi, O. Bunk, *IUCrJ* **2019**, *6*, 267.

[11] M. Liebi, V. Lutz-Bueno, M. Guizar-Sicairos, B. M. Schönbauer, J. Eichler, E. Martinelli, J. F. Löffler, A. Weinberg, H. Lichtenegger, T. A. Grünewald, *Acta Biomaterialia* **2021**, *134*, 804.

[12] F. K. Mürer, B. Chattopadhyay, A. S. Madathiparambil, K. R. Tekseth, M. Di Michiel, M. Liebi, M. B. Lilledahl, K. Olstad, D. W. Breiby, *Sci Rep* **2021**, *11*, 2144.

[13] T. A. Grünewald, M. Liebi, N. K. Wittig, A. Johannes, T. Sikjaer, L. Rejnmark, Z. Gao, M. Rosenthal, M. Guizar-Sicairos, H. Birkedal, M. Burghammer, *Sci. Adv.* **2020**, *6*, eaba4171.

[14] P. Fratzl, R. Weinkamer, *Progress in Materials Science* **2007**, *52*, 1263.

[15] P. T. Corona, B. Berke, M. Guizar-Sicairos, L. G. Leal, M. Liebi, M. E. Helgeson, *Phys. Rev. Materials* **2022**, *6*, 045603.

[16] A. Rodriguez-Palomo, V. Lutz-Bueno, M. Guizar-Sicairos, R. Kádár, M. Andersson, M. Liebi, *Additive Manufacturing* **2021**, *47*, 102289.

[17] A. Rodriguez-Palomo, V. Lutz-Bueno, X. Cao, R. Kádár, M. Andersson, M. Liebi, *Small* **2021**, *17*, 2006229.

[18] F. Schaff, M. Bech, P. Zaslansky, C. Jud, M. Liebi, M. Guizar-Sicairos, F. Pfeiffer, *Nature* **2015**, *527*, 353.

[19] M. Liebi, M. Georgiadis, A. Menzel, P. Schneider, J. Kohlbrecher, O. Bunk, M. Guizar-Sicairos, *Nature* **2015**, *527*, 349.

[20] M. Georgiadis, A. Schroeter, Z. Gao, M. Guizar-Sicairos, M. Liebi, C. Leuze, J. A. McNab, A. Balolia, J. Veraart, B. Ades-Aron, S. Kim, T. Shepherd, C. H. Lee, P. Walczak, S. Chodankar, P. DiGiacomo, G. David, M. Augath, V. Zerbi, S. Sommer, I. Rajkovic, T. Weiss, O. Bunk, L. Yang, J. Zhang, D. S. Novikov, M. Zeineh, E. Fieremans, M. Rudin, *Nat Commun* **2021**, *12*, 2941.

[21] M. Guizar-Sicairos, M. Georgiadis, M. Liebi, *J Synchrotron Rad* **2020**, *27*, 779.

[22] M. Georgiadis, A. Schroeter, Z. Gao, M. Guizar-Sicairos, D. S. Novikov, E. Fieremans, M. Rudin, *NeuroImage* **2020**, *204*, 116214.

[23] S. Waltersperger, V. Olieric, C. Pradervand, W. Glettig, M. Salathe, M. R. Fuchs, A. Curtin, X. Wang, S. Ebner, E. Panepucci, T. Weinert, C. Schulze-Briese, M. Wang, *J Synchrotron Rad* **2015**, *22*, 895.
[24] W. Glettig, D. Buntschu, E. Panepucci, M. Wang, A. Omelcenko, *Proc. 12th Int. Conf. on Mechanical Engineering Design of Synchrotron Radiation Equipment and Instrumentation (MEDSI2023)* **2023**.
[25] A. Casanas, R. Warshamanage, A. D. Finke, E. Panepucci, V. Olieric, A. Nöll, R. Tampé, S. Brandstetter, A. Förster, M. Mueller, C. Schulze-Briese, O. Bunk, M. Wang, *Acta Crystallogr D Struct Biol* **2016**, *72*, 1036.
[26] J. A. Wojdyla, E. Panepucci, I. Martiel, S. Ebner, C.-Y. Huang, M. Caffrey, O. Bunk, M. Wang, *J Appl Crystallogr* **2016**, *49*, 944.
[27] J. A. Wojdyla, J. W. Kaminski, E. Panepucci, S. Ebner, X. Wang, J. Gabadinho, M. Wang, *J Synchrotron Rad* **2018**, *25*, 293.
[28] T. C. Huang, H. Toraya, T. N. Blanton, Y. Wu, *Journal of Applied Crystallography* **1993**, *26*, 180.
[29] CXS group, PSI, **2022**.
[30] O. Bunk, M. Bech, T. H. Jensen, R. Feidenhans'l, T. Binderup, A. Menzel, F. Pfeiffer, *New J. Phys.* **2009**, *11*, 123016.
[31] A. Streun, T. Garvey, L. Rivkin, V. Schlott, T. Schmidt, P. Willmott, A. Wrulich, *J Synchrotron Rad* **2018**, *25*, 631.
[32] F. Nolting, C. Bostedt, T. Schietinger, H. Braun, *Eur. Phys. J. Plus* **2023**, *138*, 126.
[33] Andrew Allen, Jan Ilavsky, Fan Zhang, Gabrielle Long, Pete Jemian, **2009**.
[34] F. Leonarski, S. Redford, A. Mozzanica, C. Lopez-Cuenca, E. Panepucci, K. Nass, D. Ozerov, L. Vera, V. Olieric, D. Buntschu, R. Schneider, G. Tinti, E. Froejdh, K. Diederichs, O. Bunk, B. Schmitt, M. Wang, *Nature Methods* **2018**, *15*, 799.
[35] E. Fröjdh, A. Bergamaschi, B. Schmitt, *Front. Phys.* **2024**, *12*, 1304896.
[36] L. C. Nielsen, M. Liebi, P. Erhart, **2023**.
[37] L. C. Nielsen, P. Erhart, M. Guizar-Sicairos, M. Liebi, *Acta Crystallogr A Found Adv* **2023**, *79*, 515.
[38] T. Rolvien, F. N. Schmidt, P. Milovanovic, K. Jähn, C. Riedel, S. Butscheidt, K. Püschel, A. Jeschke, M. Amling, B. Busse, *Sci Rep* **2018**, *8*, 1920.
[39] C. Morris, B. Kramer, E. F. Hutchinson, *Clin. Anat.* **2018**, *31*, 1158.
[40] Y. Kuroda, K. Kawaai, N. Hatano, Y. Wu, H. Takano, A. Momose, T. Ishimoto, T. Nakano, P. Roschger, S. Blouin, K. Matsuo, *J Bone Miner Res* **2021**, *36*, 1535.
[41] L. Anschuetz, M. Demattè, A. Pica, W. Wimmer, M. Caversaccio, A. Bonnin, *Hearing Research* **2019**, *383*, 107806.
[42] W. Wagermaier, A. Gourrier, B. Aichmayer, in *Smart Materials Series* (Eds: P. Fratzl, J.W.C. Dunlop, R. Weinkamer), Royal Society Of Chemistry, Cambridge, **2013**, pp. 46–73.
[43] S. Pabisch, W. Wagermaier, T. Zander, C. Li, P. Fratzl, in *Methods in Enzymology*, Elsevier, **2013**, pp. 391–413.
[44] P. Fratzl, M. Groschner, G. Vogl, H. Plenk, J. Eschberger, N. Fratzl-Zelman, K. Koller, K. Klaushofer, *J Bone Miner Res* **2009**, *7*, 329.

**Supporting Information**

Fast Small-Angle X-ray Scattering Tensor Tomography: An Outlook into Future Applications in Life Sciences

*Christian Appel[a], Margaux Schmeltz[a], Irene Rodriguez-Fernandez[a], Lukas Anschuetz[d], Leonard C. Nielsen[b], Ezequiel Panepucci[a], Tomislav Marijolovic[a], Klaus Wakonig[a], Aleksandra Ivanovic[acd], Anne Bonnin[a], Filip Leonarski[a], Justyna Wojdyla[a], Takashi Tomizaki[a], Manuel Guizar-Sicairos[af], Kate Smith[a], John H. Beale[a], Wayne Glettig[a], Katherine McAuley[a], Oliver Bunk[a], Meitian Wang[a]\* and Marianne Liebi[abe]\**

a Photon Science Division, Paul Scherrer Institute, Villigen, Switzerland

b Department of Physics, Chalmers University of Technology, Gothenburg, Sweden

c ARTORG Center for Biomedical Engineering Research, Universität Bern, Bern, Switzerland

d Department of Otorhinolaryngology, Head and Neck Surgery, Inselspital, University Hospital and University of Bern, Bern, Switzerland

e Institute of Materials, École Polytechnique Fédérale de Lausanne (EPFL), Lausanne, Switzerland

f Institute of Physics, École Polytechnique Fédérale de Lausanne (EPFL), Lausanne, Switzerland

**Reconstructions of the malleus**

As mentioned in the main article, reconstructions of the malleus bone were performed on 150 projections, measured within 1.2 hours in total. We used a q range of 0.17…0.22 nm$^{-1}$ which is still representative for the mineral particle scattering, but sufficiently low in q to ensure sufficient good photon statistics as pointed out in the analysis of the main article. For details, please check out the analysis performed in figure 2. The quality of data for the illustrated bone allows us to identify similar channels as for the incus as well as collagen the main orientation of the scattering which we can associate with the collagen fibril orientations. Q-resolved reconstructions were not performed.

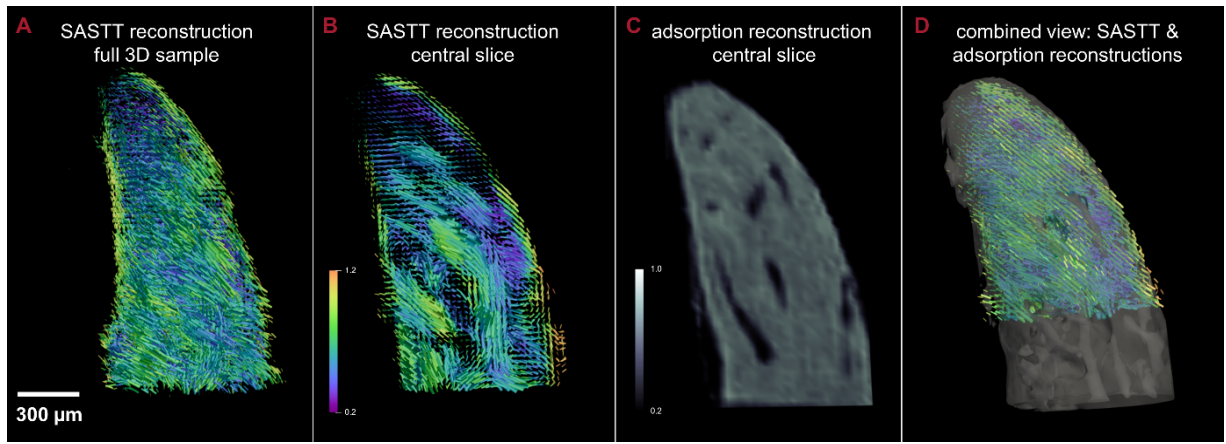

*Figure 6: SAS-TT reconstructions from 150 projections of the malleus bone, acquired at a frame rate of 500 Hz within 1.2 hrs in total. The same 3D glyph presentation is used as in Figure 5 of the main manuscript. With (A) full reconstructed volume, (B) central cut through the SAS-TT reconstructions, (C) absorption data for the same cut and the combined view of isosurface rendering together with 3D glyph data in (D).*

**TOC Figure**

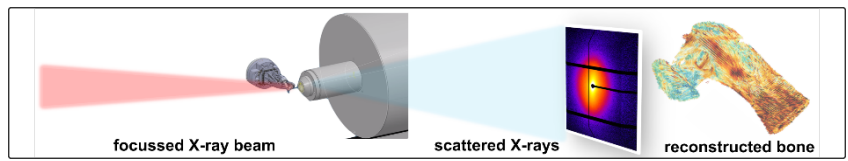

*This study reveals the future potential of SAXS tensor tomography at macromolecular X-ray crystallography beamlines and Life Sciences applications. A new standard in fast data acquisition has been set, allowing for up-to 10x higher throughput. The measurements show mineralised collagen fibrils within the incus bone of the human auditory ossicle.*